\documentclass[nohyperref]{article}

\usepackage{microtype}
\usepackage{graphicx}
\usepackage{subfigure}
\usepackage{booktabs} 

\usepackage{hyperref}

\usepackage[accepted]{ai4science}

\usepackage{amsmath}
\usepackage{amssymb}
\usepackage{mathtools}
\usepackage{amsthm}
\usepackage{comment}
\usepackage{xcolor}
\usepackage{ulem}

\usepackage{url}

\usepackage[capitalize,noabbrev]{cleveref}

\theoremstyle{plain}

\theoremstyle{definition}

\theoremstyle{remark}

\usepackage[textsize=tiny]{todonotes}
\icmltitlerunning{}

\newcommand{\particlenumber}{N}
\newcommand{\systemsize}{L}
\newcommand{\ket}[1]{$\vert #1 \rangle$}
\newcommand{\numnodefeat}{N_{f}}

\begin{document}

\twocolumn[
\icmltitle{Towards Learning Self-Organized Criticality of Rydberg Atoms using Graph Neural Networks}



\icmlsetsymbol{doneAtMxm}{$\dagger$}

\begin{icmlauthorlist}
\icmlauthor{Simon Ohler}{TUK,MXM,doneAtMxm}
\icmlauthor{Daniel Brady}{TUK}
\icmlauthor{Winfried Lötzsch}{MXM}
\icmlauthor{Michael Fleischhauer}{TUK}
\icmlauthor{Johannes S. Otterbach}{MXM}
\end{icmlauthorlist}

\icmlaffiliation{TUK}{Department of Physics, University of Kaiserslautern, Kaiserslautern, Germany}
\icmlaffiliation{MXM}{Merantix Momentum, AI Campus Berlin, Germany}

\icmlcorrespondingauthor{Simon Ohler}{sohler@rhrk.uni-kl.de}
\icmlcorrespondingauthor{Johannes Otterbach}{johannes.otterbach@merantix.com}

\icmlkeywords{Graph Neural Networks, AI4Science, Physics Simulation, Rydberg Atoms}

\vskip 0.3in
]



\printAffiliationsAndNotice{\workDoneAtMxm} 

\begin{abstract}
Self-Organized Criticality (SOC) is a ubiquitous dynamical phenomenon believed to be responsible for the emergence of universal scale-invariant behavior in many, seemingly unrelated systems, such as forest fires, virus spreading or atomic excitation dynamics. SOC describes the buildup of large-scale and long-range spatio-temporal correlations as a result of only local interactions and dissipation. The simulation of SOC dynamics is typically based on Monte-Carlo (MC) methods, which are however numerically expensive and do not scale beyond certain system sizes. We investigate the use of Graph Neural Networks (GNNs) as an effective surrogate model to learn the dynamics operator for a paradigmatic SOC system, inspired by an experimentally accessible physics example: driven Rydberg atoms. To this end, we generalize existing GNN simulation approaches to predict dynamics for the internal state of the node. We show that we can accurately reproduce the MC dynamics as well as generalize along the two important axes of particle number and particle density. This paves the way to model much larger systems beyond the limits of traditional MC methods. While the exact system is inspired by the dynamics of Rydberg atoms, the approach is quite general and can readily be applied to other systems.
\end{abstract}

\begin{figure}
    \centering
    \includegraphics[width=0.48\textwidth]{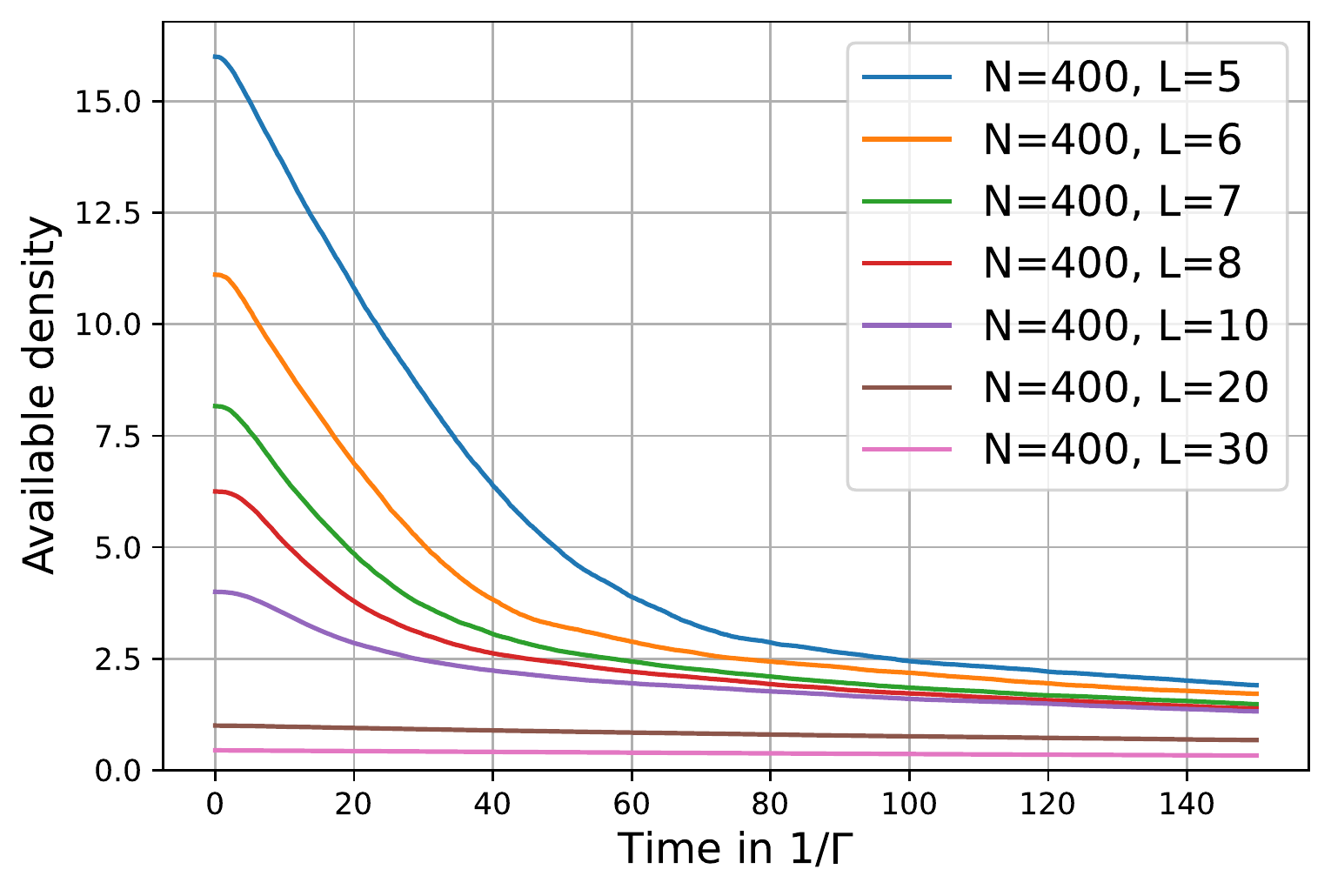}
    \caption{Results from the $60000$-timestep rollout of the GNN, after being trained only on single timestep data. The available density (sum of ground and active state atom densities) shows the characteristic SOC behavior: all initial densities above the critical value of $n_{c}\approx 1$ converge to the same point, whereas all curves that start below $n_{c}$ show only a slow decay.}
    \label{fig:decay_crit_density}
\end{figure}

\section{Introduction}
Deep Learning methods deliver impact for many applications ranging from Computer Vision, Natural Language Processing, Audio and Speech Processing to Optimal Control. In recent years we also see these methods drive innovation in areas of scientific interest, such as drug discovery \cite{JimenezLuna2020DrugDW}, protein folding \cite{Jumper2021HighlyAP}, molecule dynamics \cite{No2020MachineLF, Gilmer2017NeuralMP,Henderson2021ImprovingMG} as well as the dynamics of classic multi-particle systems \cite{Battaglia2016InteractionNF, SanchezGonzalez2020LearningTS} and many more. In the context of particle dynamics the class of Graph Neural Networks (GNNs) \cite{Bronstein2021GeometricDL, SanchezGonzalez2020LearningTS, Shlomi2021GraphNN, Lemos2022RediscoveringOM} are particularly successful. We build up on this success and investigate the application of GNNs to learn the time-evolution of a complex many-body system in the regime of so-called self-organized criticality (SOC).

SOC represents a widespread phenomenon observed in many different fields, ranging from noise in electric circuits \cite{Bak1987SOC}, earthquakes \cite{Sornette1989SelfOrganizedCA}, forest fires \cite{Malamud1998ForestFA, Drossel1992SelforganizedCF} over the spread of information or diseases \cite{Gleeson2014MemePop, Rhodes1996PowerLG} to atomic excitation dynamics \cite{Helmrich2020SignaturesOS}. The term refers to the property of a complex system to drive itself towards the critical point between two dynamical phases in the relevant parameter space during its time evolution without any external control or fine-tuning. Remarkably, even though the microscopic processes in these systems may be very different, certain properties such as the scaling laws of the critical behaviour remain the same, a phenomenon termed \textit{universality}. This allows to study the properties of SOC dynamics in one system, e.g. forest fires, through controlled simulation and experimentation of other systems, e.g. atoms. An important challenge in studying SOC-behavior is the efficient simulation of the dynamics of large systems. State-of-the-art methods employ Markov-Chain Monte-Carlo (MCMC) methods that are limited in terms of number of particles and system size \cite{wintermantel2021epidemic}. This renders the study of large-scale systems, required for reliable predictions of key SOC properties such as critical exponents, hard. Additionally, the simulation requirements in time and memory increase with the system size and make the simulations prohibitively expensive in the absence of high-performance computation infrastructure.

In this study, we investigate the use of GNNs to accurately approximate the time-evolution operator of a complex system inspired by the dynamics of a classical ensemble of Rydberg atoms in the SOC regime. The GNN acts as an effective surrogate model and replaces the costly MCMC step operator.
We model the atoms as nodes in a GNN with internal degrees of freedom encoded in the node attribute and the Euclidean distances between the endpoints attached as edge attributes. We learn the one-timestep update function of the node attributes from data simulated using classical MCMC methods. We demonstrate that the GNN is capable to efficiently learn this operator and reliably reproduce the characteristic properties of the system such as the facilitation dynamics, Rydberg blockade effects and SOC excitation trajectories. Moreover, we show early indications that the model generalizes over several orders of magnitudes along two important axes: particle number and system size (particle density). This property enables the application of this model to system sizes beyond what can be simulated using classical MCMC methods, while also enabling computations on moderate hardware.

In addition, we are, to the best of our knowledge, the first to extend the simulation capabilities of GNNs to systems with dynamics in internal as well as external degrees of freedom. While previous work was mostly focused on simulating the position dynamics of the nodes \cite{Battaglia2016InteractionNF, kipf2018neural, SanchezGonzalez2020LearningTS}, an external degree of freedom, we also discuss the dynamics of the node attribute, i.e. the internal degree of freedom, based on the interaction of a node with its neighbors.

We make both the code for our experiments as well as the Monte Carlo code to create our datasets publicly available upon acceptance.

\section{Related Work}
GNNs have been used in a multitude of applications. Most relevant to our study is the work by Sanchez-Gonzalez et al.~\yrcite{SanchezGonzalez2020LearningTS} to learn the dynamics of challenging physics motions such as viscous fluids and elastic bodies simulated using third party tools. They do so by breaking the objects into a discrete set of volumes that are subsequently represented as nodes in a graph. The position update of the volume is learned by predicting the next position using an encoder-decoder system with several message passing (MP) layers. 

Pfaff et al.~\yrcite{Pfaff2021LearningMS} expand on this work to simulate the dynamics of wings and flags under external forces. They do so by introducing two different meshes, one for the surface of the object under study and a second one encoding the real-world, Euclidean distance between the nodes of the surface mesh. Using MP they learn to predict the surface mesh positions in the next time-step from simulated data.

Battaglia et al.~\yrcite{Battaglia2016InteractionNF} study the use of graph representations in the data to enable a multi-layer perceptron to reason about complex interactions of physical systems. By making the relationships between physical objects explicit in the graph data structure, they are able to predict the orbital motion of planets as well as the dynamics of solids under non-trivial, external constraints.

Shlomi et al.~\yrcite{Shlomi2021GraphNN} give a review of the use of GNNs in high-energy particle physics. They draw attention to the benefits of using GNNs to model the different particle interactions through clever design of the node, edge and vertex attributes when representing the data as relationship graphs.

Lemos et al.~\yrcite{Lemos2022RediscoveringOM} use GNN simulators in conjunction with symbolic regression to rediscover Newton's law of gravity from simulated data of multi-planetary motions. This serves as a demonstration to infer physical laws from complex dynamics as multi-planetary motion can behave chaotically under certain circumstances.

Martinkus et al.~\yrcite{Martinkus2021ScalableGN} discuss the construction of a hierarchical GNN in order to scale the simulation of particle dynamics from a few hundreds to tens of thousands. By recursively partitioning the space and compute interactions between the cells at each hierarchy they are able to reduce the $\mathcal{O}(N^2)$ interaction computation to $\mathcal{O}(N \log N)$ complexity.

Prakash \& Tucker~\yrcite{Prakash2021GraphNF} investigate the use of multi-task learning in GNNs to simultaneously learn the dynamics and the unknown physical parameters of an ensemble of objects with Newtonian dynamics. They show that using this approach one can circumvent the design of inductive biases in the GNN construction, making the GNN architecture more versatile across datasets.

Olsson \& Noé~\yrcite{olsson2019dynamic} use a graphical model to describe molecular kinetics, where each molecular subsystem changes its internal state based on interactions with the neighboring subsystems. Instead of using a single global state, the approach of using interacting subsystems is key in reducing the computational effort.

Jin \& Voth~\yrcite{jin2018ultra} study interfacial systems using an ultra-coarse grained model that is constructed by including the local particle density in the internal states of the coarse-grained sites. This construction allows to distinguish different phases and to successfully capture phase coexistence.

Zhang et al. \yrcite{zhang2022dynamic} use a GNN architecture not for physical simulations, but for sequential recommender systems in social networks. Instead of using only the previous user-item interactions of a single person to predict future interactions, the authors include dynamic collaborative signals among different users.

In contrast to these works, we are the first to discuss the dynamics of a complex atomic many-body system whose entities have an additional internal state, i.e. where the node attributes can change depending on the neighbors. This extends the use of GNNs to efficiently simulate systems that have internal as well as external degrees of freedom. On top of this we also demonstrate the ability of the GNNs to learn non-trivial spatio-temporal correlations, despite being trained on a spatially local, one-step prediction task.

\begin{figure*}[t!]
    \centering
    \subfigure[]{
    \includegraphics[width=0.2\textwidth]{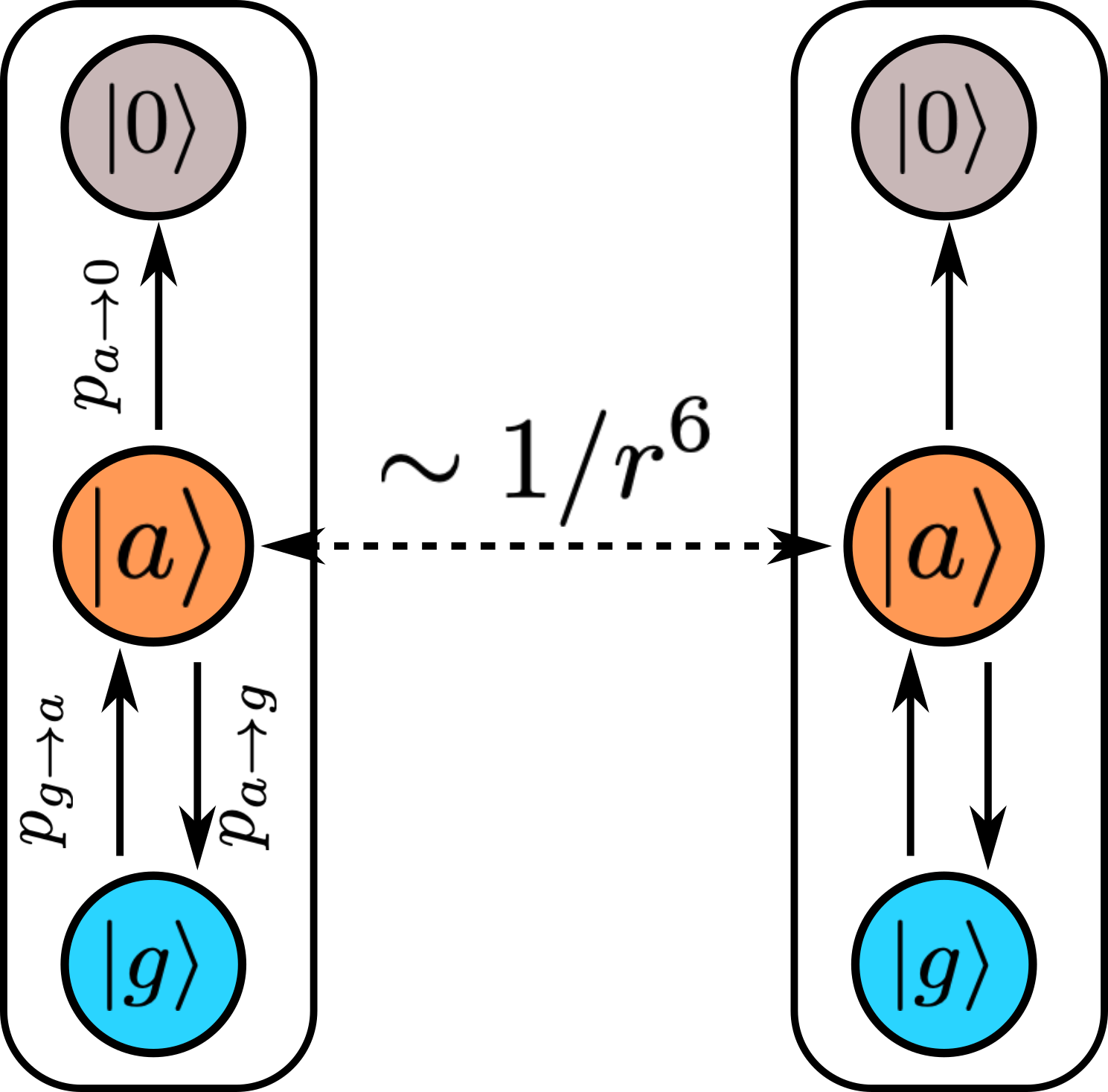}
    \label{fig:transition_rates}}
    \hspace{3cm}
    \subfigure[]{
    \includegraphics[width=0.45\textwidth]{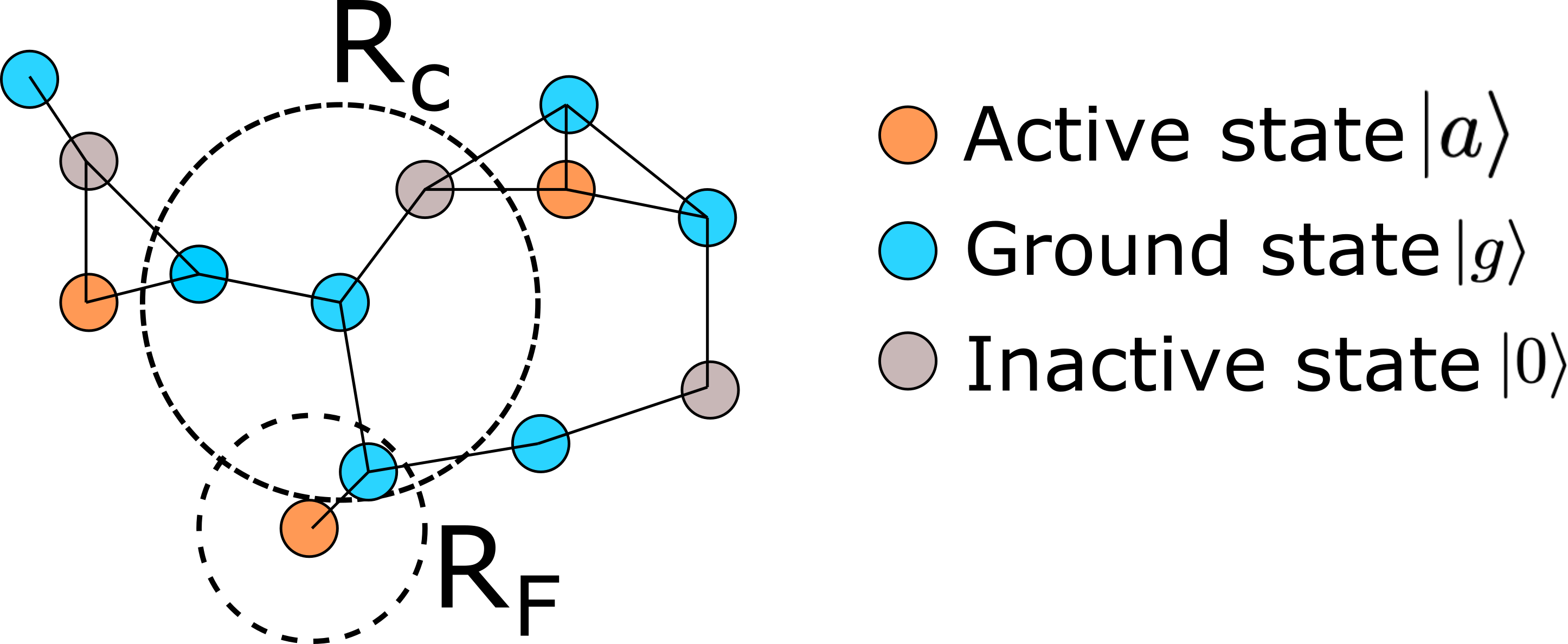}
    \label{fig:gas_atoms}}
    \caption{Schematics for the SOC problem construction. Figure~\ref{fig:transition_rates}: The model entities (atoms) have each three internal states. The ground state \ket{g}, the active state \ket{a} and the inactive state \ket{0}. The atoms have a distance-dependent interaction scaling as $1/r^6$. Figure~\ref{fig:gas_atoms}:  When constructing the input graph there are two length scales that are relevant: the facilitation radius $R_F$ denoting the relevant interaction distances and the cutoff radius $R_C$ used to limit the number of edges in the graph to avoid all-to-all connectivity.}
    \label{fig:my_label}
\end{figure*}

\section{Self-Organized Criticality in Rydberg Atoms}
A typical SOC system consists of many identical entities, each taking one of three internal states. The entity's state can vary over time as a function of the system's dynamics. Since in our case the entities are Rydberg atoms, we label these three states as the \textit{ground}, \textit{active} and \textit{inactive} states, respectively (see Figure~\ref{fig:gas_atoms}). For illustration, in disease spread models these states are commonly referred to as susceptible, infected and recovered. The key ingredient to produce the characteristic SOC dynamics lies in the transitions between the internal states of each atom, as shown in Figure \ref{fig:transition_rates}: the ground state \ket{g} can only become active via the neighborhood-dependent rate $p_{g\rightarrow a}(\particlenumber)$, called the  \textit{facilitation} process. The active state \ket{a} can undergo the reverse process and return to the ground state again with a neighborhood-dependent rate $p_{a \rightarrow g}(\particlenumber)$, or transition into the inactive state \ket{0}, which is a terminal state in the process.

The physical realization of the SOC system we study here is based on a gas of Rydberg atoms subject to external laser fields. The microscopic transition rates for each atom $i$ per time interval $\delta t$ of the model are given by
\begin{align}
    p_{i, g \rightarrow a} &= 1 - \textrm{e}^{-R_{i}\delta t}\label{eqn:excitation}
    \\
    p_{i, a \rightarrow g} &= (1-\epsilon_{i}) \left(1 - \textrm{e}^{-(R_{i}+\gamma)\delta t} \right)\label{eqn:deexcitation}
    \\
    p_{i, a \rightarrow 0} &= \epsilon_{i} \left(1 - \textrm{e}^{-(R_{i}+\gamma)\delta t} \right)
    \\
    R_{i} & =  \frac{2\Omega^{2}\gamma}{\gamma^2 + \Delta_{i}^{2}}\label{eqn:r_s_rate}
    \\
    \epsilon_{i} & = \frac{b}{1+R_{i} / \Gamma}
    \\
    \Delta_{i} & = \Delta_0 \left(-1 + \sum_{\ell \in N} \frac{1}{r_{i,\ell}^6}\kappa_{\ell}\right) \label{eq:resonance}
    \\
    \kappa_{\ell} &=
    \begin{cases}
			1, & \text{if $s_{\ell}=\vert a\rangle$ }\\
            0, & \text{otherwise}
	\end{cases}
\end{align}
The parameters $\Delta_0, \Gamma, b, \gamma, \textrm{ and } \Omega$ are details of the system and can be treated as hyper-parameters to the problem. $\particlenumber$ denotes the total number of atoms in the system. The symbol $\kappa_{\ell}$ in Equation \eqref{eq:resonance} ensures that the sum includes only the active atoms (state $s_{\ell}=\vert a\rangle$). Additionally, $r_{i, \ell}$ denotes the Euclidean distance between atom $i$ and $\ell$ in natural units, where we set the length scale $R_{F}=1$ (see Figure \ref{fig:gas_atoms}). A more detailed introduction to this system is presented in Appendix \ref{app:RydbergStateSystem}. Equations \eqref{eqn:excitation} to \eqref{eq:resonance} describe a strongly interacting many-body system, for which neither trivial nor analytic solutions exist. In order to solve the system for large particle numbers and long times, approximate techniques such as MCMC \cite{wintermantel2021epidemic} or coarse-graining  \cite{Helmrich2020SignaturesOS} have historically been used. It is worth pointing out that Equation~(\ref{eq:resonance}) displays a resonance behavior dependent on the number and distance of the neighboring active atoms, giving rise to the neighborhood-dependent state transition rates: If $|\Delta|$ is large, the transition rate is suppressed and no change in the dynamics is happening. If an active atom is present at the right distance to an atom $i$, then $\Delta_{i}=0$ and the transition is facilitated. The facilitated atom can in turn facilitate the excitation of another atom. Above a critical density this leads to an explosive spread of excitations. The system hence shows two distinct phases:
\begin{itemize}
    \item The \textit{active} phase is characterized by a large number of active atoms that spread through the entire system. The density of inactive atoms is small, and each active atom facilitates on average more or equal to one ground state atom before it ceases to be active.
    \item The \textit{absorbing} phase is a state where the available density of ground state atoms is so low that transitions into the active state typically become inactive quickly before they can spread, i.e. facilitate the transition of another ground state atom into the active state.
\end{itemize}
The critical point of the system occurs at the point where the two phases meet, i.e. where an active atom facilitates exactly one ground state atom on average before it decays. Since the inactive state is a dead-end, the available density of ground and active state atoms can only decrease over time. Consequently, the active phase automatically moves towards the critical point, which constitutes a self-organization of the critical phase termed SOC. Since SOC systems at the critical point share certain universal properties, being able to simulate a single SOC system with high accuracy and scalability allows investigations of many different physical systems.
\section{Learning the Rydberg SOC Dynamics with a GNN}
\paragraph{Training data.}
We model the atoms as the nodes of a GNN with the internal states of the atoms mapped to node attributes and the distance between them as the edge attributes (see Figure \ref{fig:gas_atoms}). An advantage of the GNN structure over conventional discretization schemes is the independence of system density from the discretization length and thus an easier computational description of the problem at hand. Note that throughout this study we always use training data for two-dimensions, where all atoms are placed into a square box in the $xy$-plane.

The training data is generated using classical Monte-Carlo methods (see Appendix \ref{app:MCMC}) and we create a large dataset covering a broad range of particle numbers and densities to expose the model to a diverse set of scenarios during training. Each training sample consists of $X$, a matrix of shape $\left(\particlenumber, \numnodefeat\right)$, and the ground truth $Y$, a matrix of shape $\left(\particlenumber, 3\right)$.
Here, $\numnodefeat$ denotes the number of node features, which are the position and internal state of each node. Note that $X$ is not directly the input to the GNN, since we replace the absolute positions with relative distances. The ground truth matrix $Y_{i}=(P_i(g\vert s_{i}), P_i(a\vert s_{i}), P_i(0\vert s_{i}) )$ contains for each node $i$ with the internal state $s_{i}$ the probabilities to change to or remain in each of the three states in the next timestep. Before feeding a graph into the GNN, we apply the following physically motivated processing steps to facilitate easier learning for the model:
\begin{itemize}
    \item We mask out all nodes in the inactive state \ket{0}, since they do not contribute to the dynamics and can be neglected without loss of accuracy.
    \item The edges between nodes are created by a radius-graph scheme: For two nodes to be connected their relative (Euclidean) distance needs to be smaller than a critical radius $R_c$, which is a hyperparameter for our model. Increasing $R_c$ yields higher accuracy, but also increases the computational burden due to a quadratic rise in the number of neighbors. However, as the interaction drops off as $1/r^6$ the benefit of a large $R_c$ is minimal. For all our calculations we use $R_c=2$ (in natural units).
    \item Two nodes in the ground state \ket{g} do not interact with each other, allowing us to remove edges between them. This causes no loss of accuracy and reduces the total number of edges in a graph by up to $80\%$, since the fraction of active atoms is typically lower than $15\%$.
    \item Each edge is given the Euclidean distance between its nodes as an edge attribute. The absolute position data is not encoded to encourage translation and rotational invariance of the model, since only relative distances are relevant for the SOC behavior.
\end{itemize}

\paragraph{Network architecture.}
To construct the network architecture we first note some characteristics of the probability distribution for the state transition probabilities. As the data is generated using Monte-Carlo simulations of a continuous process, the state transition probabilities are small in order to minimize the numerical error of the simulation. As a result, the probability of each atom to remain in its current state is large in order to preserve probability. This results in a bimodal distribution of the state transition probabilities with the actual transitions orders of magnitude smaller, i.e. $\mathcal{O}(10^{-7} - 10^{-1})$ than the probabilities to remain in the current state  $\mathcal{O}(1)$ (For more details see Appendix \ref{app:MCMC}). However, the dynamics of the system are encoded in the changes in the state transition probabilities and without accounting for the sizable differences in the probabilities the model will simply learn the large background probabilities and not pay much attention to the subtle fluctuations in the transition probabilities that are caused by neighborhood interaction effects.

\begin{figure}
    \centering
    \includegraphics[width=0.25\textwidth]{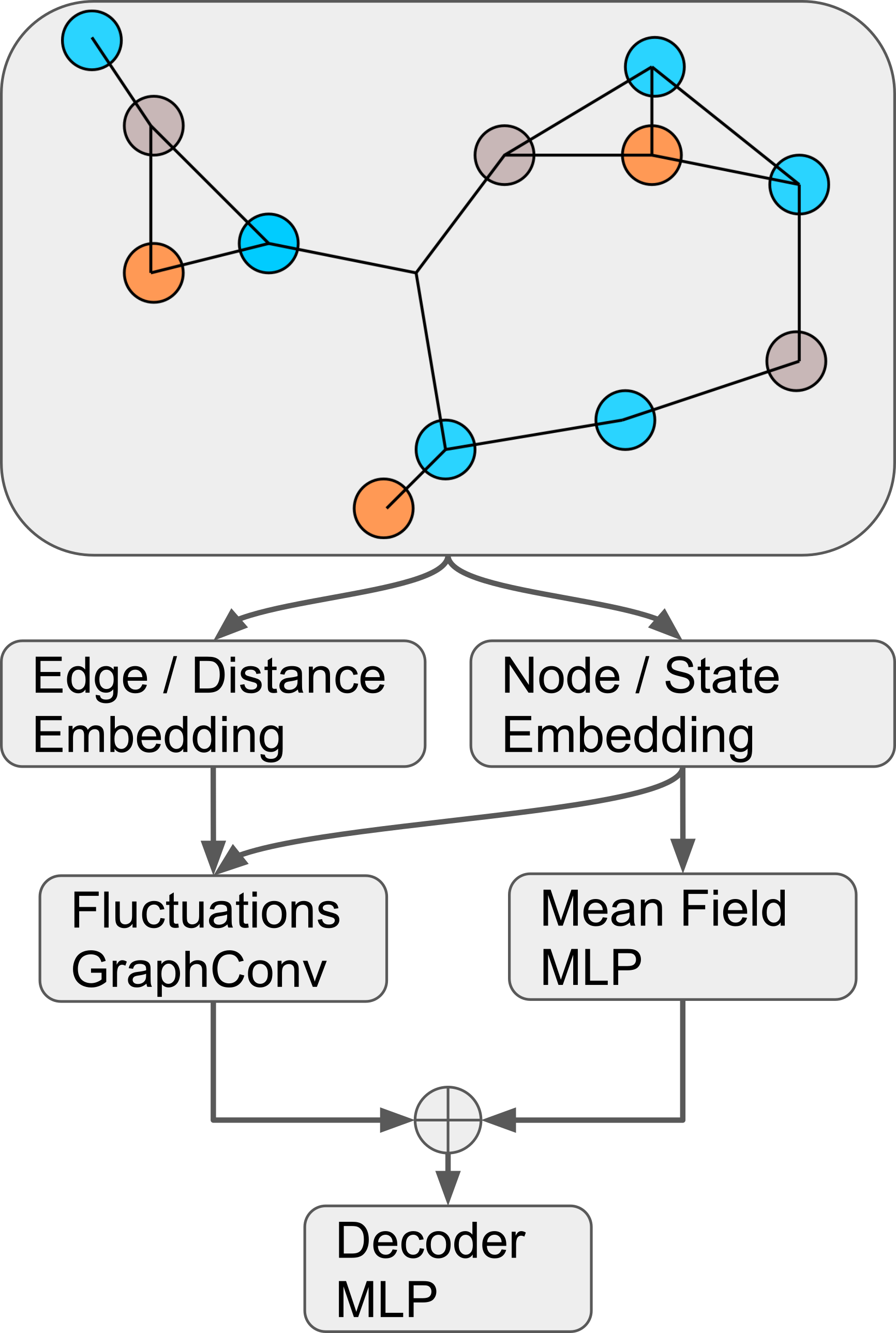}
    \caption{We use a residual architecture for the network structure to encourage the model to learn the background mean-field and the fluctuations on top of the mean field separately. To this end we feed the node embeddings into a simple MLP that learns the factorized node distribution. The fluctuations are learned using a Graph Convolution network that receives the node as well as edge embeddings as input. Finally, we add the mean-field and fluctuations before feeding them through the final MLP to predict the update probabilities.}
    \label{fig:network}
\end{figure}

To address this challenge, we make use of a residual network architecture as depicted in Figure~\ref{fig:network}, where we give the network the opportunity to model the large background distribution, termed the \textit{mean-field} contribution, and the neighborhood effects of the changes in the state transition probabilities, termed \textit{fluctuations}. This terminology is in accordance with corresponding techniques in the traditional modelling of dynamic systems, where one tries to describe the system in terms of small variations, the fluctuations, around a stationary point, the mean-field.
More specifically, we embed the state of each node into a higher-dimensional embedding space. Note that we removed the inactive state \ket{0} from the data. We then feed each node state individually through a simple feed forward network. Since this network does not take into account the edge attribute, it effectively learns only the factorized, non-correlated features of the nodes, which corresponds to the mean-field part of the dynamics. To capture the correlations, we combine the embedded node states and edges and feed them through a Message Passing Graph Convolution layer \cite{Gilmer2017NeuralMP}. We then add the outputs of the mean-field and fluctuation modules before passing them through a final MLP network to predict the state transition probabilities of each node. We use \textsc{ReLU} nonlinearities as activation funtions and normalize the activations using \textsc{LayerNorm} for the MLPs.

\paragraph{Loss function.}
As noted before the transition probability distribution changes over many orders of magnitude and hence we need a loss function that works over such a large interval. We experimented with mean-squared-error (MSE) and Kullback-Leibler divergence (KLD) and found the latter to be more stable. This can be motivated by the fact that the logarithmic dependency of the KLD penalizes differences in the exponents of the probabilities much more severely than the MSE loss that ignores such differences if the probabilities are small in favor of minimizing the differences for large probabilities.

\section{Experiments \& Results}
Having established the modelling approach, we now turn to the experimental validation.  We implement the GNN based on \textsc{PyTorch} \cite{Pytorch2019} using the \textsc{PyTorch Lightning} framework \cite{PytorchLightning} as well as \textsc{PyTorch Geometric} \cite{PyTorchGeometric2019}. The network consists of a $2D$ embedding module for the internal node state with an output dimension of $10$, a state encoder network (mean-field component) consisting of $2$ fully connected layers with hidden dimension of $50$ and a final output dimension of $32$, a Graph Convolution layer with a single message passing step and a final decoder with 2 fully connected layers and hidden dimension $50$. The Graph Convolution uses the \textsc{NNConv} module from \cite{Gilmer2017NeuralMP} that is implemented in \textsc{PyTorch Geometric} and has an output dimension of $64$. The MLP inside of the \textsc{NNConv} has a single hidden layer with $50$ units. Inside the fluctuations component we apply another MLP of $2$ hidden layers with $100$ units each and an output dimension of $32$, before adding both components. In total the network has approximately $81000$ parameters.

For all experiments we use the Adam optimizer \cite{Kingma2015AdamAM} with a learning rate of $10^{-3}$, beta values of $(0.9, 0.999)$ and batch size $15$. We set the cutoff radius $R_c=2$ (in natural units) and train the model for $25$ epochs with $6000$ training instances each on an Intel® Xeon® Gold Processor 6154 with a peak memory usage of about $240$MB. The training instances are drawn from three datasets that we created that all contain $N\approx 150$ atoms per instance but had varying densities. In two datasets we used $dx=0.25$ and $dx=0.5$, respectively. Here, we define $dx$ as $dx = \systemsize / \sqrt{\particlenumber}$,
where $\systemsize$ is the length of one edge of the square box in which we place the atoms. Intuitively, $dx$ represents the distance between the atoms if they were placed in an ideal square lattice. In the third we artificially increased the fraction of active state atoms to $50\%$ at a density of $dx=1.25$ to address the fact that active atoms are otherwise rare in the training data. Finally, for reasons of numerical stability we transform the probabilities using a square-root transformation. This ensures the probabilities stay in the range $[0,1]$ while mitigating the extreme bimodal nature of the ground-truth probability. See the Appendix \ref{app:MCMC} for more details.

\begin{figure*}
    \centering
    \subfigure[Model prediction of the facilitation process of a ground state atom as well as the ground-truth curve from Monte Carlo.]{
    \includegraphics[width=0.45\textwidth]{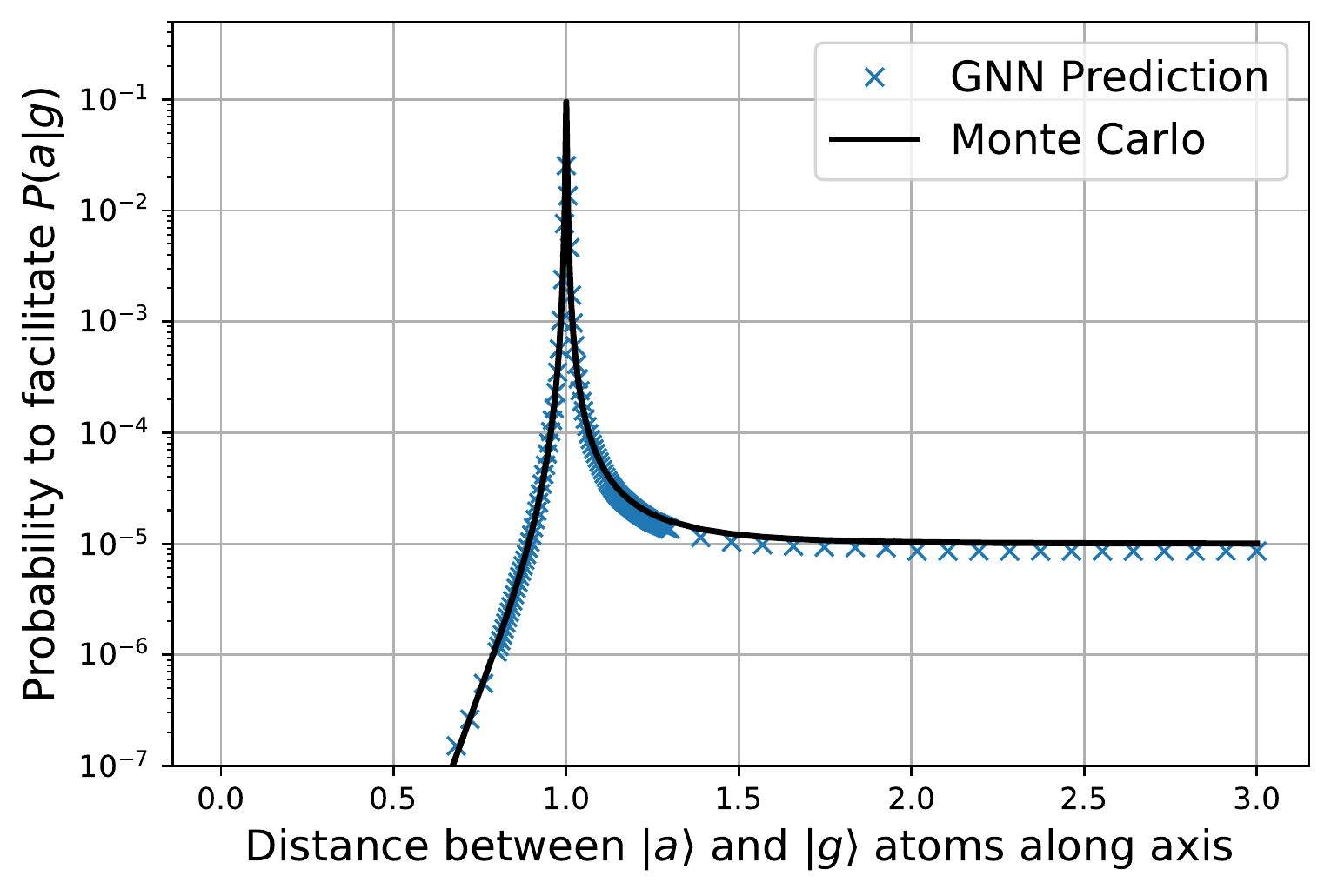}
    \label{fig:pred_facil}
    }
    \subfigure[Model prediction of the facilitated de-excitation of an active atom as well as the ground-truth curve from Monte Carlo. The inset shows a magnification of the peak around $r=1$.]{
    \includegraphics[width=0.45\textwidth]{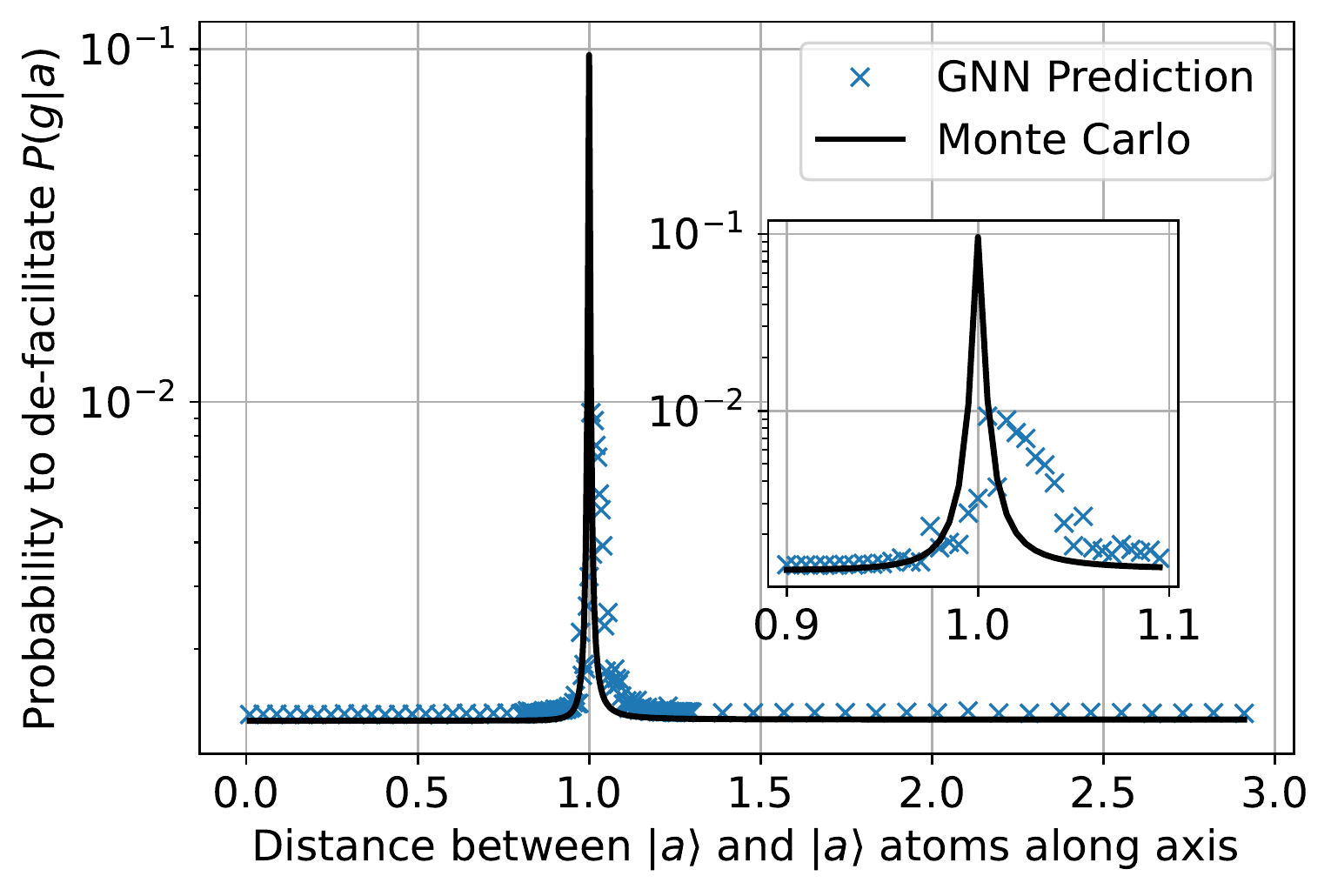}
    \label{fig:pred_defacil}}
    \caption{Benchmark and validation curves to assess correct learning of key interaction features for the SOC dynamics in Rydberg atoms.}
\end{figure*}

\paragraph{Transition rates of internal node dynamics.}
To the best of our knowledge there is no comparable publication that predicts transition probabilities in SOC systems or considers the dynamics of internal degrees of freedom of a node. Hence we developed metrics to assess the quality and validate the learning of the GNN. We base these metrics on key aspects of the Rydberg-SOC dynamics.

The fastest process in the SOC system is the \textit{facilitation} process by which an active atom promotes surrounding ground state atoms to the active state, according to the resonance feature of Equation (\ref{eq:resonance}). The facilitation probability of a ground state atom at a distance $r$ to an active atom is characterized by an extremely sharp peak at $r=1$ (in natural units), several orders of magnitudes larger than the non-interacting ($r\to\infty$) case. Additionally, as $r\to 0$ the facilitation probability decreases rapidly to zero, due to the \textit{Rydberg blockade} \cite{lukin2001dipole} (also see \cite{saffman2010quantum} for a review). For $r\gg R_F$, the facilitation probability tends to a small, but finite value. To check this behaviour we construct a custom graph where we place an active state atom at the origin and a ground state atom at the position $\left(x, y\right) = \left(r, 0\right)$. Additionally, we randomly place other ground state atoms in the graph to keep the total particle number and density close to those seen during training. The value we record is then $P_i(a\vert g)$ of the ground state atom to become active at position $\left(x, y\right) = \left(r, 0\right)$. We then sweep this setting across different values of $r$, creating a new random graph each time. The result of this experiment is shown in Figure~\ref{fig:pred_facil} and shows excellent agreement with the exact solution of the facilitation rate around the facilitation peak $r=1$ (in natural units) and in the $r\to\infty$ limit. In the blockade regime for $r\ll 1$ the GNNs prediction converges to $P_i(a\vert g)\sim 10^{-11}$, six orders of magnitude smaller than the limit $r\to\infty$. The true convergence to zero however is not captured. This is an expected result given the fact that even making large relative errors on very small ground-truth probabilities incurs a small loss penalty for the algorithm. Nonetheless, since the predicted blockaded facilitation probability is sufficiently small compared to all SOC timescales, we hypothesize that this deviation from the ground truth does not lead to qualitative changes in the GNNs predictions.

\begin{table}[!t]
    \centering
    \begin{tabular}{lll}
        Task & Configuration & KLD \\ \hline
        num. particles & N=50, dx=0.5, d=2 & 1.17\textrm{e}-4 \\ 
        & N=150, dx=0.5, d=2 & 1.50\textrm{e}-4 \\ 
        & N=500, dx=0.5, d=2 & 1.83\textrm{e}-4 \\ 
        & N=1000, dx=0.5, d=2 & 1.97\textrm{e}-4\\
        & N=1500, dx=0.5, d=2 & 2.04\textrm{e}-4\\
        \hline
        density & N=150, dx=0.125, d=2 & 3.00\textrm{e}-5 \\ 
        & N=150, dx=0.25, d=2 & 8.17\textrm{e}-5 \\ 
        & N=150, dx=0.5, d=2 & 1.50\textrm{e}-4 \\ 
        & N=150, dx=1, d=2 & 7.54\textrm{e}-5\\
        & N=150, dx=2.5, d=2 & 1.99\textrm{e}-5\\
    \end{tabular}
    \caption{Test loss averaged over $2000$ graphs each from different datasets, where either the particle number or density was varied. For the case of particle number generalization, we test on up to $10$ times as many particles as seen during training and observe that the loss rises slowly with the particle number. For the case of varying densities we go a factor of $2$ smaller than the smallest density seen during training (including the engineered dataset with more active atoms), as well as a factor of two above. In both cases, the loss is reduced compared to $dx=0.5$, which we attribute to less facilitation events and more blockade (for higher density, low $dx$) or more non-interacting cases (for lower density, high $dx$).
    }
    \label{tbl:generalization}
\end{table}

The inverse process is called \textit{de-facilitation} and is, to the best of our knowledge, a process unique to Rydberg atoms. In this case, a de-excitation (return to the ground state) of an active atom is facilitated by the presence of a second active atom. If these two active atoms are close to each other, the probability of one atom to return to the ground state increases sharply at the relative distance $r=1$. Even though both the facilitation and the de-facilitation share the same physical origin, their ground-truth curves in Figures \ref{fig:pred_facil} and \ref{fig:pred_defacil} look dissimilar on short distances. The is due to the second process that allows active atoms to return to the ground state, which is independent of the neighborhood (see $\gamma$ in eqn. \eqref{eqn:deexcitation}). We check this behaviour the same way as for the facilitation metric, but in this case the atom at position $\left(x, y\right) = \left(r, 0\right)$ is also active and we record $P_i(g\vert a)$. The results are shown in Figure~\ref{fig:pred_defacil}. The de-facilitation is much harder to learn as the case of two active atoms at close distance is very rare in the dataset, especially in the already very narrow region around $r=1$ where the behavior changes drastically. We find that the GNN accurately predicts the non-resonant case as well as the position of the peak ar $r=1$. However, the interval between $r\in\left[0.995, 1.005\right]$, where the ground truth values increase by a factor of $10$, is not reproduced, as the GNNs prediction does not exceed $0.009$, an order of magnitude short of the ground truth.

\paragraph{Particle number and density generalization.}
In addition to the SOC specific metrics we investigate the prediction accuracy on graphs consisting of more nodes, i.e. higher particle number, or denser, i.e. shorter average distance between particles, than those seen during training. The results are summarized in Table~\ref{tbl:generalization}. We see that when trained on $N\approx 150$ atoms, the model suffers no significant loss of accuracy when predicting on a varying number of atoms up to $N=1500$. The lowest test loss occurs at smaller $N$ than during training, which might be the result of boundary effects that the model learns when trained on small system sizes. This scalability of the GNN has been observed in other particle-based regression tasks \cite{SanchezGonzalez2020LearningTS} and is most likely due to the local scope of the message passing, which encodes spatial translation invariance of the dynamics operator.

In addition, we see that the test loss of the GNN  predicting on datasets with different densities is also small across a range of average $dx$ distances.  During training, the model is confronted with densities $dx=0.5$ and $dx=0.25$ as well as artificial graphs with $dx=1.25$. The test case of $dx=0.5$ on an unseen dataset shows the highest test loss, which is most likely due to the fact that more facilitation events happen at this density than at lower or higher densities. In the case of lower densities, the chance of a ground state atom to be located in the small interval around $r=1$ to an active atom is reduced, and in the case of larger densities the case of $r\ll 1$ occurs more frequently, where we see extremely small state-change probabilities (blockade).

\begin{figure}[h!]
    \centering
    \includegraphics[width=0.45\textwidth]{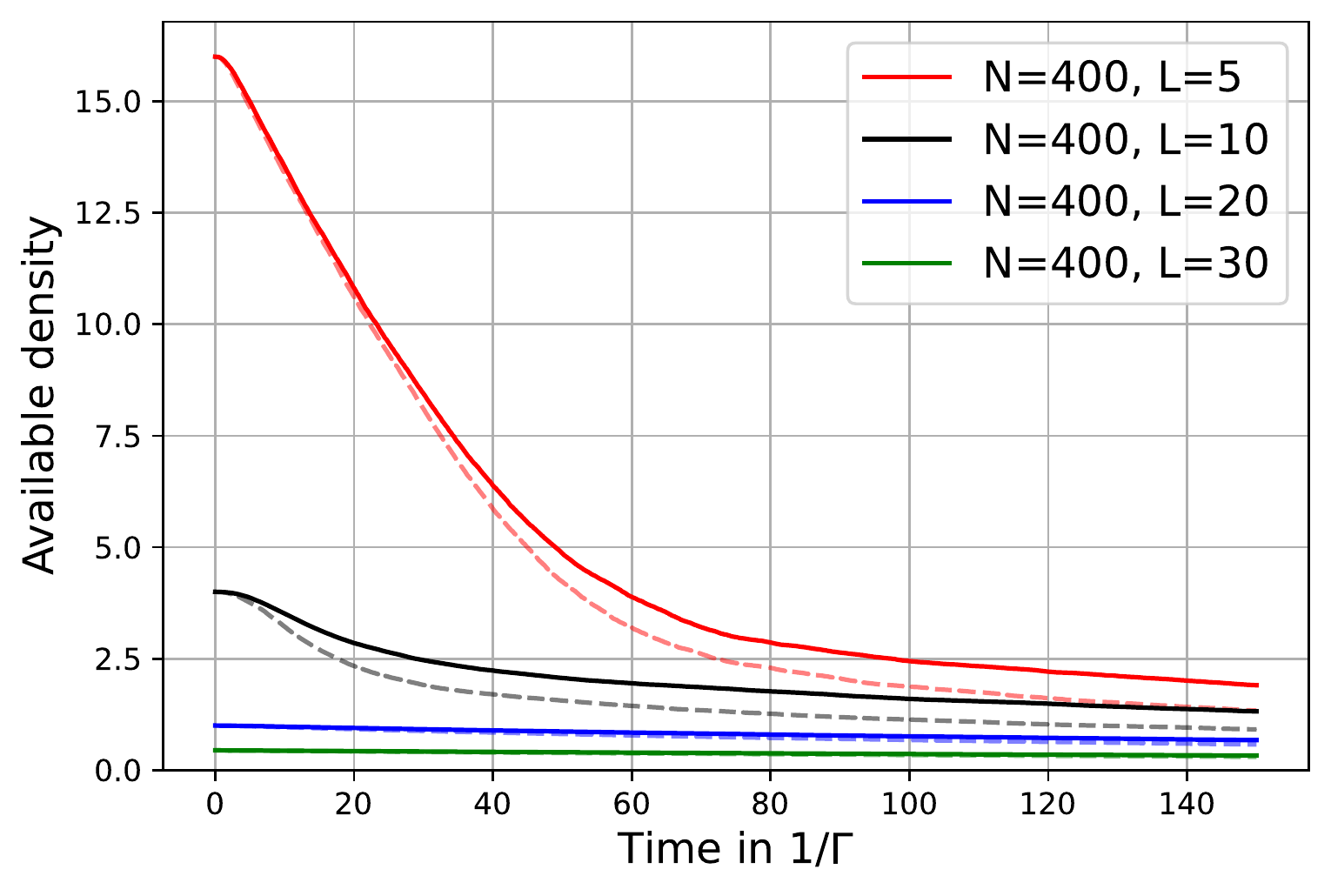}
    \caption{Comparison between the $60000$-timestep predictions of the GNN (solid) and the Monte-Carlo simulation (dashed) from which the training data was obtained. We do not show all initial conditions in the interest of readability. The available density is the sum of ground and active state densities.}
    \label{fig:trajectory_comparison}
\end{figure}

\paragraph{SOC excitation trajectories.}
Having established that the model captures key aspects of the SOC dynamics and is able to generalize over particle numbers as well as densities, we can use the trained model to iteratively predict the time evolution of an initial ensemble of atoms. To this end we randomly create an initial 2D distribution of atoms, each in the ground state. Additionally, we assign a velocity to each atom which we sample from a Maxwell-Boltzmann distribution with the most probable velocity $\hat{v}=1$. We input this initial state into the graph network and use the output probabilities to roll new internal states for each node and construct a new graph, where we record the fraction of atoms not in the inactive state. Then, we input the new graph into the GNN and repeat this procedure for $60,000$ steps. In each step, we update the positions of each node using the velocity: $\Vec{x}\rightarrow\Vec{x}+\Vec{v}\delta t$. We use repelling boundary conditions, where the corresponding velocity component is reflected when a particle collides with the boundary. Lastly we average the resulting excitation trajectories of $50$ runs with equivalent initial conditions $(\particlenumber, \systemsize)$. The number of steps is chosen such that the total predicted time $T$ is equivalent to $T=60,000\,\delta t=150\Gamma^{-1}$ since it takes approximately $T=150\Gamma^{-1}$ for the system to reach the critical state. Here $\Gamma=1$ (the decay rate of the active state) serves as the natural unit of (inverse) time in the system. The results are shown in Figure~\ref{fig:decay_crit_density}. All curves that have an initial density above the critical density $n_{c}\approx 1$ (in the active phase) feature three distinct regimes: (i) a fast initial decay, (ii) a convergence to the same critical density $n_{c}$ and (iii) a slow decay beyond that. All curves that start below the critical density (absorbing phase) exhibit only feature (iii). This is a hallmark characteristic of SOC behavior. When comparing the GNN predictions to the MCMC results in Figure \ref{fig:trajectory_comparison} we see that the GNN shows a slower decay in phase (i). This effect is less pronounced the lower the initial density, which suggests the deviation stems from imperfect reproduction of the facilitation peak in Figure \ref{fig:pred_facil}. Indeed, since the GNN underestimates the transition probability very close to $r=1$, less facilitation events occur here than in the Monte Carlo baseline, causing a slower initial spread of excitations.

\paragraph{Ablation study of the GNN architecture.}
We chose a residual architecture for the GNN model. The two branches can be interpreted as the mean-field and the fluctuations branch. The former is a simple MLP and only receives as input the node embeddings, while the latter contains a graph convolutional layer and is supplied with both the node and edge embeddings. Since the fluctuations branch alone could learn the MCMC update function, we perform an ablation study to test the performance changes of the model when switching off the mean-field branch of the architecture.

Using the identical training procedure as for the full model, we find that the test loss of the non-residual model is increased by a factor of two. The performance on the facilitation metric shown in Figure \ref{fig:pred_facil} remains unaffected; however learning the de-facilitation (\ref{fig:pred_defacil}) is significantly worse. The de-facilitation peak at $r=1$ is predicted at $r=0.975$ (full model: $r=1.005$) at a transition probability of $P(g|a)=0.0026$, which is $36\times$ smaller than the correct result (full model: $10\times$). Additionally, the training of the non-residual model appears significantly less stable as indicated by noisier training curves, including frequent large-magnitude jumps, which are absent in the training of the full model. In this sense, the residual architecture in form of a mean-field branch, although not strictly necessary, represents an important ingredient to obtain high-quality predictions.

\paragraph{Discussion of the results.}
There are a few key implications we want to discuss based on the experimental results. We showed that the GNN learns the one-step time-evolution operator successfully and can display the hallmark characteristics of SOC dynamics. However, by construction, the GNN convolution operators are spatially translation invariant and local in nature. This implies that the long-range correlation of the SOC phase requires a time-dimension and is truly spatio-temporal as the spatial locality of the Graph Convolution does not allow to learn long-range spatial correlations.

A second aspect is that the spatial translation invariance of the Graph Convolution operator allows us to scale the time-evolution of the system to very large particle numbers, as all that matters is the local connectivity of the graph, not the overall particle number. Hence, the traditional $\mathcal{O}(N^2)$ scaling for MCMC does not apply.

Finally, the use of a graph data structure enables us to also scale over many densities, if the model has learned the facilitation metrics correctly. This is in contrast to standard simulations based on the discretization of the real-space coordinate. Due to the necessity of having a fine-grained grid, the simulated densities are typically small in that case in order to approximate continous systems. However, this limitation does not exist for GNNs as the relevant scale is the distance encoded in the edge attribute $r$ as a fraction of the natural distance $R_F$. Taken together with the particle number scaling, it might allow the study of dense systems with large particle numbers.

\section{Summary}
We present a new method to model the time dynamics of SOC systems using machine learning on graphs. We use a physically motivated, lightweight GNN architecture to learn the one-step time-evolution operator of each atom of the SOC system. In contrast to previous studies, we extend the GNN simulation toolkit to learn the dynamics of internal degrees of freedom, i.e. the node-state, in addition to the external degrees of freedom, i.e. the distance between atoms. We demonstrate high accuracy of the predictions via excellent agreement of the learned facilitation rate compared to exact ground-truth rates. Additionally, we show that the model generalizes to at least one order of magnitude more particles than seen during training without significant loss of accuracy, as well as higher and lower densities than it was trained on. Moreover, we showed that the model can successfully reproduce the decay to a common critical density and the separation of time-scales in the excitation trajectories, a hallmark characteristic of SOC, despite having been trained on single-timestep data only.

We leave the extraction of critical exponents and other SOC elements as well as scaling to even larger particle numbers to future work. Additionally, our model could most likely benefit from hierarchical graph methods as presented in \cite{Martinkus2021ScalableGN}, which could reduce the computational complexity of larger systems. Although we see that the graph creation and distance calculation scales linearly for our GNN compared to the quadratic scaling of all-to-all Monte-Carlo algorithms, a quantitative analysis of the computational cost and runtime of our model compared to the currently used MCMC approaches is left to future work. Lastly, the Monte Carlo training data that we used could be replaced by real observations of SOC systems. In that case, the GNN model could learn directly the interaction laws and circumvent the time discretization of differential equations, thereby providing an independent check of the underlying theory.

\textbf{Contributions.} SO designed, implemented and ran the experiments using GNNs and drove the project development. DB implemented the classic MCMC methods to generate the data. WL supported the GNN network design and technical implementation as well as the training deployment. MF designed and supervised the project. JSO designed and supervised the project and experiments and supported the implementation. SO, DB and MF thank the DFG for their generous support under SFB TR 185, Project
Number 277625399. The simulations were executed on the high performance cluster "Elwetritsch" at the TU Kaiserslautern which is part of the "Alliance of High Performance Computing Rheinland-Pfalz" (AHRP). We kindly acknowledge the support of the RHRK.
\bibliography{social}
\bibliographystyle{icml2022}

\newpage
\appendix
\onecolumn
\section{Derivation of Rate Equations from Optical Bloch Equations / State-transition description of a Driven Rydberg Gas}\label{app:RydbergStateSystem}
In this section, we shortly summarize the derivation of the rate equations discussed in the main text. To help readers draw parallels with the physics literature, we will use the subscript $r$ for the active (rather: Rydberg) state and the subscript $g$ for the ground state. The derivation considers the density matrix $\rho$ for a two-level system consisting of the ground and Rydberg states that is coupled to an external light field which we assume to be classical (not in the single-photon limit). Since we take the full quantum mechanics of the two-level atom into account, this is called the \textit{semi-classical} approach. For a two level system with atom-light interaction in the semi-classical limit, the optical Bloch equations are then given as \cite{scully_quantum_1997}
\begin{align}
    \dot{\rho}_{rr} &= -i\Omega (\rho_{rg} - \rho_{gr}) - \Gamma \rho_{rr}, \\
    \dot{\rho}_{rg} &= -\gamma \rho_{rg} + i\Delta \rho_{rg} - i\Omega (\rho_{rr} - \rho_{gg}),
    \intertext{where $\rho_{ij}$ represents the entries of the density matrix. For large dephasing $\gamma$, $\dot{\rho}_{rg}$ can be adiabatically eliminated. This gives}
    \rho_{rg} &= \frac{i\Omega (\rho_{rr} - \rho_{gg})}{-\gamma + i\Delta} \\
    \dot{\rho}_{rr} &= -\frac{2\Omega^2 \gamma}{\gamma^2 + \Delta^2} (\rho_{rr} - \rho_{gg}) - \Gamma \rho_{rr} \\
    &\equiv -\Gamma_\uparrow \rho_{rr} + \Gamma_\downarrow \rho_{gg}.
\end{align}
This gives the transition rates $\Gamma_\uparrow$ from the ground to excited state and $\Gamma_\downarrow$ from the excited to the ground state. To simulate these rates, we choose the microscopic parameters in such a way as to achieve a sufficient separation of timescales, see table \ref{tbl:parameter_values}.
\begin{table}[h]
    \centering
    \begin{tabular}{ll}
        parameter & value \\ \hline
        $\Delta/\Gamma$ & $2000$ \\
        $\Omega/\Gamma$ & $20$ \\
        $\Gamma$ & $1$ (Unit of time) \\
        $\gamma/\Gamma$ & $20$ \\
        $\delta t\Gamma$ & $0.0025$ \\
        $b$ & $0.5$ \\
    \end{tabular}
    \caption{Microscopic parameters used in the creation of Monte-Carlo data. We use units where $\hbar=1$.}
    \label{tbl:parameter_values}
\end{table}
Using these values, we find the following separation of timescales:
\begin{align}
    \Gamma_\text{facil}&=\frac{2\Omega^{2}}{\gamma}=40\Gamma
    \\
    b\Gamma&=0.5\Gamma,
    \\
    \Gamma_\text{non-int}&=\frac{2\Omega^{2}\gamma}{\gamma^{2}+\Delta_{0}^{2}}\approx 0.004\Gamma,
    \\
    \Gamma_{\text{facil}}&=80\,b\Gamma
    \\
    b\Gamma&\approx 125 \Gamma_{\text{non-int}}.
\end{align}
To observe SOC behavior, a separation of approximately a factor of $100$ between each rate is recommended.
\section{Monte-Carlo simulation of the Training Data}
\label{app:MCMC}
\subsection{Creation of Training Data}
As outlined in the main text, the training data for the GNN is generated using classical Monte-Carlo methods where each training instance consists of $X$, a matrix of shape $\left(\particlenumber, \numnodefeat\right)$, and the ground truth $Y$, a matrix of shape $\left(\particlenumber, 3\right)$. $Y_{i,s}=(P_i(g\vert s_{i}), P_i(a\vert s_{i}), P_i(0\vert s_{i}) )$ contains for each node $i$ with the internal state $s_{i}$ the probabilities to change to or remain in each of the three states in the next timestep.

In each training instance, even though the total number of particles is fixed for every dataset, the distribution of the three internal node states may be different. Since over the course of a SOC simulation (see Figure \ref{fig:decay_crit_density}) the fractions of ground, active and inactive states change drastically, we set bounds for the fraction of each state and draw instances at random. These bounds are
\begin{align}
    p(g)&=\left[0.15, 1.0\right]
    \\
    p(a)&=\left[0, 0.15\right]
    \\
    p(0)&=\left[0, 0.9\right].
\end{align}
Since the \ket{0} atoms are masked out at every step, this leads to a varying particle number for each training instance.
\begin{figure}
    \centering
    \includegraphics[width=0.85\textwidth]{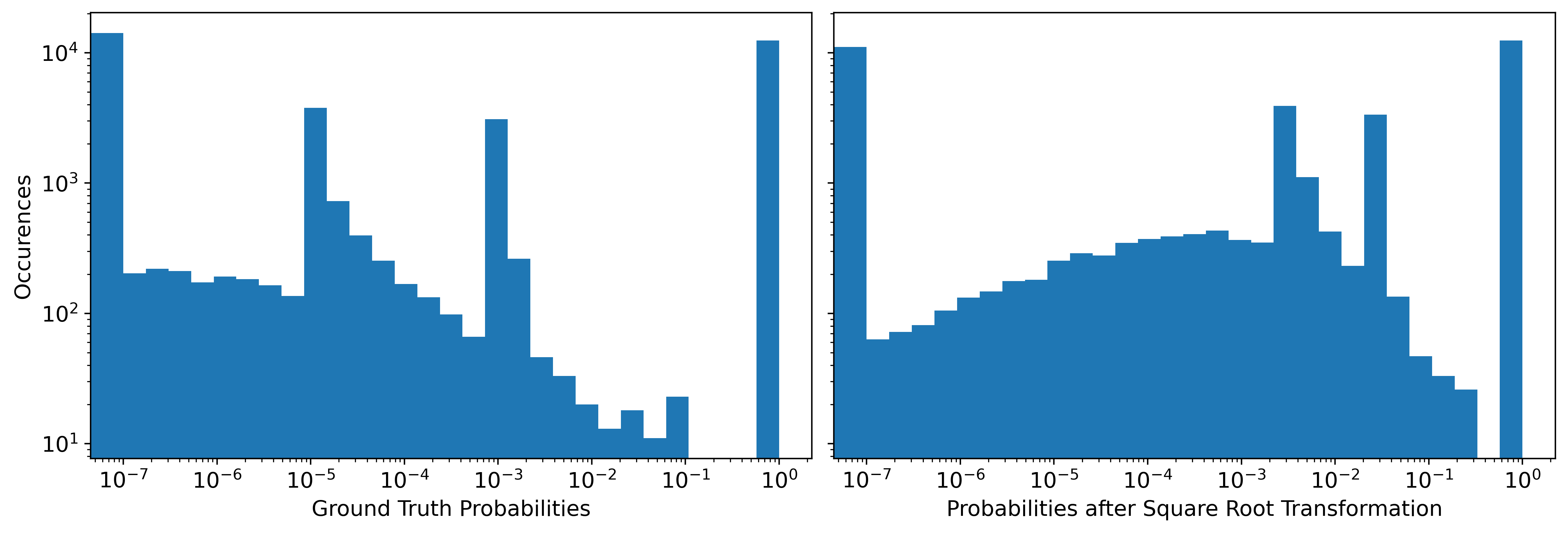}
    \caption{Left: Histogram over the target probabilities. The leftmost bin contains all values smaller than $10^{-7}$. One notices the peaks at $0.9$ and higher and at around $10^{-3}$. Right: Histogram of the ground truth probabilities after the square-root transformation.}
    \label{fig:probab_distribution_target}
\end{figure}

\subsection{Square Root Transformation}
In defining a model architecture, we are guided by the expected distribution of transition probabilities $P(i\vert s)$, where $i, s\in\{\text{ground}, \text{active}, \text{inactive}\}$ states. The training data stems from fixed-timestep Monte Carlo, where the timestep has to be chosen to be small against the inverse rate at which the states of the atoms change, such that the error stemming from time discretization is small. As a result, the probability of a node to remain in its current state is always large, independent of the states of neighboring nodes. While neighborhood-effects can increase the probability of state-change by factors of ten-thousand, the absolute probability of any state-change is nonetheless $<10\%$. Additionally, the SOC system imposes further constraints on the transition probabilities. To observe critical behaviour in the SOC regime, the three timescales in the system need to be separated by at least a factor of $100$ each (see Appendix \ref{app:RydbergStateSystem}).
Since the time increment $\delta t$ must be chosen sufficiently short such that the transition probability connected to the largest rate is about $10^{-1}$, the transition probability connected to the smallest rate can maximally be $10^{-5}$ (see Figure \ref{fig:pred_facil} for $r\to\infty$). Lastly, to accurately capture the Rydberg blockade effect for small inter-atomic distances (see Figure \ref{fig:pred_facil} for $r\to 0$), we need to model transition probabilities to at least two orders of magnitudes below the slowest transition rate at maximally $10^{-7}$. This results in a bi-modal target distribution, where most values are in the interval $\left[0.9, 1.0\right]$ as well as in the interval $\left[10^{-4}, 0.1\right]$, including a long tail for even smaller probabilities as shown in \ref{fig:probab_distribution_target}. Learning probabilities as small as $10^{-7}$ is very challenging, as noise during training becomes significant for such small numbers. To avoid this, we decided to perform a transformation on the ground truth data that increases the magnitude of very small numbers but keeps large ones effectively unchanged, while still producing numbers in the interval $\left[0, 1\right]$ as required by the loss function. For this purpose we chose the square root function, which we apply node-wise on all three probabilities before normalizing their sum to unity. To obtain real predictions from the model, it is then necessary to take the normalized square from the model's output.

We found that this transformation generally increased training stability as well as performance on the (de-)facilitation metrics.


\end{document}